\documentclass[twocolumn,prl]{revtex4}
\usepackage{epsfig}
\usepackage{dcolumn}
\usepackage{bm}
\usepackage{amssymb}

\def\be{\begin{equation}}
\def\ee{\end{equation}}

\def\lan{\langle}
\def\ran{\rangle}

\begin{document}

\title{Colloidal gelation, percolation and structural arrest}
\author{A. de Candia $^{a,b}$, E. Del Gado$^{a,b}$,
A. Fierro $^{a,c}$, N. Sator $^{d}$, and A. Coniglio $^{a,c}$}
\affiliation{${}^a$ Dipartimento di Scienze Fisiche and INFN Sezione di Napoli,
Universit\`a di Napoli ``Federico II'', via Cintia 80126 Napoli, Italy}
\affiliation{${}^b$ CNISM Napoli}
\affiliation{${}^c$ Coherentia CNR-INFM}
\affiliation{${}^d$ Laboratoire de Physique Th\'eorique des Liquides
Universit\'e Pierre et Marie Curie
UMR CNRS 7600 Case 121, 4 Place Jussieu 75252 Paris Cedex 05, France}
\date{\today}
\begin{abstract}
By means of molecular dynamics, we study a model system for colloidal
suspensions where the interaction is based on a competition between
attraction and repulsion.
At low temperatures the relaxation time first increases as a power law
as a function of the volume fraction 
and then, due to the finite lifetime of the bonded structures, it deviates
from this critical behavior.
We show that colloidal gelation at low temperatures and low volume
fractions is crucially related to the formation of spanning long living
cluster, as we suggested in the last few years.
Besides agreeing with experimental findings in
different colloidal systems, our results shed new light on the
different role played by the formation of long living bonds
and the crowding of the particles in colloidal structural arrest.
\end{abstract}
\maketitle
In attractive colloids a rich phenomenology is found in the temperature -
volume fraction plane \cite{edinburgh}-\cite{attr_exp}. At high
temperature a hard sphere glass transition occurs
at a volume fraction, $\phi\simeq 0.57$.
By decreasing the temperature, at high volume fraction, the effect of the short
range attraction produces an attractive glass line with a
reentrant line well described in the framework of the mode
coupling theory \cite{attr,attr_sim}, and confirmed by experiments.
At low temperature and increasing the volume fraction a much more complex
situation arises \cite{weitzclu},
\cite{gron}-\cite{sciortino3}, characterized by
a cluster phase followed by a kinetic arrest with viscoelastic
properties very similar to those found in polymer gelation.

One of the important problems is to understand the mechanism which
gives rise to colloidal gelation. When phase separation does not
occur, colloidal gelation has been interpreted in terms of cluster
mode coupling theory \cite{kroy} or Wigner glass \cite{sciortino2}.
In contrast to these approaches, we have suggested that at low
volume fraction and low temperature, percolation phenomena play a
crucial role in colloidal gelation characterized by the presence of
a spanning cluster of bonded particles \cite{noi1,noi3} as also
found in recent experiments \cite{bartlett}. Of course, at high
enough temperature and high volume fraction, one expects a
structural arrest due essentially to the hard sphere part of the
potential. At intermediate values of temperature and volume
fractions, molecular dynamical studies have shown the occurrence of
colloidal gelation phenomena characterized by a structural arrest,
on average well described by the mode coupling theory of attractive
glasses \cite{cates}. Interestingly, the system is dynamically heterogeneous:
The set
of all particles can be partitioned in two subsets made of slow and
fast particles and structural arrest occurs close to the onset of a
spanning cluster of only the subset of slow particles \cite{cates}. One may ask
the question whether the same scenario for colloidal gelation 
can be smoothly extended 
from this intermediate regime to the low
volume fraction region.  In this letter, by using a molecular
dynamics simulations of a DLVO-type potential \cite{dlvo} for
charged colloidal systems, we give evidence of a clear crossover
from the low volume fraction regime to the intermediate regime. In
the first regime we find the bond lifetime much larger than the
structural relaxation time. As a consequence, dynamically, the clusters
behave as made of permanent bonds as in chemical gelation, and the
regime is dominated by the dynamics of such clusters. Increasing
$\phi$, when these two time scales become comparable, the structural
relaxation begins to be affected also by the crowding of the
particles, and a clear crossover to a new glassy regime is found.
Interestingly enough such crossover has also been found in some
micellar systems at rather high volume fraction
\cite{mallamaceetal,noi1}, suggesting a common mechanism
in both systems.


%

We consider a system of $N=10000\,\phi$ particles, with $\phi=0.08$--$0.25$,
interacting via a DLVO-type potential,
which contains a Van der Waals type interaction plus an effective
repulsion due to the presence of charges:
\begin{equation}
V(r)=\epsilon \left[a_1 \left(\frac{\sigma}{r}\right)^{36}
-a_2\left(\frac{\sigma}{r}\right)^6+a_3e^{-\lambda(\frac{r}{\sigma}-1)}\right],
\label{potential}
\end{equation}
where $a_1=2.3$, $a_2=6$, $a_3=3.5$, and $\lambda =2.5$ \cite{note_yukawa}.
With these parameters
the repulsion term dominates the Van der Waals attraction at long
range, providing a short range attraction and a long range repulsive barrier.
The effective repulsion in the potential prevents the liquid-gas phase
separation and stabilize the size of the clusters, as expected
\cite{gron}.
The potential is truncated and shifted at a distance of 3.5$\sigma$.
To mimic the colloidal dynamics, we performed
molecular dynamics simulations at constant temperature.
Equations of motion were solved in the canonical ensemble (with a
Nos\'e-Hoover thermostat) using a velocity Verlet algorithm \cite{nose}
with a time step of $0.001 t_0$ (where $t_0=\sqrt{\frac{m
\sigma^2}{\epsilon}}$ and $m$ is the mass of the particles).
We equilibrate the system at temperatures $k_B T=~0.2,~0.23,~0.25\epsilon$ and 
$\phi$ increasing from $0.07$ to $0.23$ \cite{lamellar}.

Here we analyze the structure and the dynamics by varying the
temperature and volume fraction. We calculate first the static structure
factor, $S(k)$, and the cluster size distribution, $n(s)$, where two particles
are linked if their relative distance is
smaller than the local maximum of the potential \cite{notahill}.

\begin{figure}[ht]
\begin{center}
\mbox{ \epsfysize=6cm\epsfbox{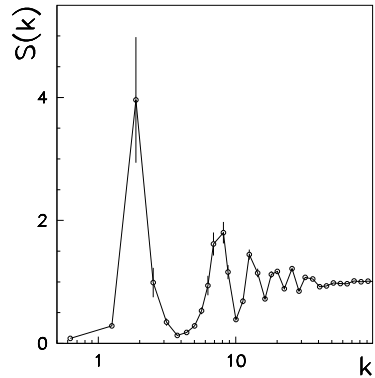} }
\end{center}
\caption{The static structure factor, $S(k)$,
at $k_BT=0.2\epsilon$ and $\phi=0.13$, displays a peak around
$k_0 \simeq2$.}
\label{figura1}
\end{figure}

\begin{figure}[ht]
\begin{center}
\mbox{ \epsfysize=6cm\epsfbox{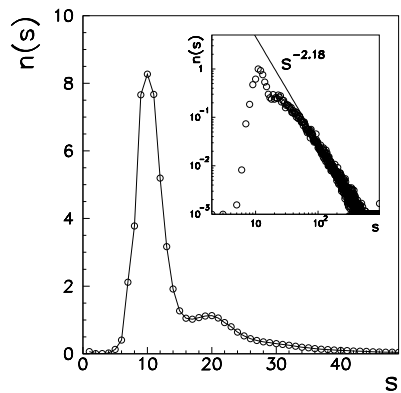} }
\end{center}
\caption{{\bf Main frame}: the cluster size distribution, $n(s)$,
at $k_BT=0.2\epsilon$ and $\phi=0.13$, displays a peak around an optimum
cluster size. {\bf Inset}:
At $\phi =0.16\simeq \phi_p$, a power law tail appears.}
\label{figura2}
\end{figure}

\begin{figure}[ht]
\begin{center}
\mbox{ \epsfysize=6cm\epsfbox{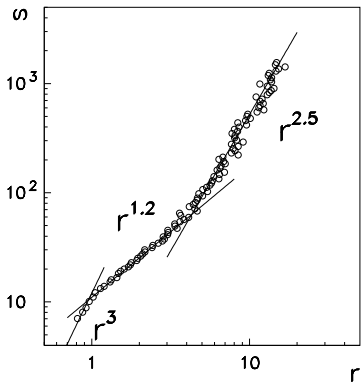} }
\end{center}
\caption{The cluster size, $s$, as a function of the radius,
$r$, at $k_BT=0.2\epsilon$ and $\phi =0.16\simeq \phi_p$:
While at very small scales clusters are compact,
on intermediate length scales the dependence suggests the presence of very thin
and
chain structures made up by clusters of typical size $s\lesssim 10$;
It crosses to a random percolation type of structure
over larger length scales.}
\label{figura3}
\end{figure}

At low temperatures, i.e. when the kinetic energy becomes smaller than the
repulsive barrier, and low volume fractions,
the static structure factor, $S(k)$, displays
a peak around $k_0 \simeq2$ (see Fig.~\ref{figura1}).
This feature is due to the fact that
the competing short range attraction and long range repulsion produce stable
clusters of a typical size.
The presence of a typical size is also manifested in the cluster size
distribution, $n(s)$, plotted in Fig.~\ref{figura2}
at $k_BT=0.2\epsilon$ and $\phi=0.13$.
It is interesting to observe the secondary
peak due to the fusion of two stable clusters.
By increasing the volume fraction $\phi$, a spanning cluster
appears at $\phi=0.16$ for $k_BT=0.2\epsilon$, and the cluster
size distribution typically displays a power law decay with an exponent
$\tau \simeq2.18$, consistent with the random percolation model \cite{staul}.
In Fig.~\ref{figura3} we show the dependence of the cluster size, $s$, on its
radius, $r$, at $\phi_p$: The data correspond to a compact structure for
clusters of dimension $s\lesssim 10$ and radius $r\lesssim 1$,
a fractal dimensionality $d_f\simeq1.2$ on intermediate length scales, and a
crossover to $d_f\simeq2.5$ at larger length scales.
These results suggest that at low volume fraction compact stable clusters form
with typical size $s\simeq10$ and radius $r\simeq1$. By increasing the volume
fraction a residual attractive interaction between the clusters produces
tube-like structures with fractal dimension $d_f\simeq1.2$ up a size $s\sim
60$. By further increasing the volume fraction these structures coalesce to
build a random percolating network. For the temperature here studied the
percolation threshold $\phi_p$ is weakly dependent on the temperature,
$0.16\lesssim\phi_p\lesssim 0.17$.


A preliminary study of this model was published in Ref.\ \cite{noi3} and these
results were presented
in the meeting ``Dynamical arrest'', hold in Bad Gastein in January 2005.
This analysis has been lately developed in Ref.\ \cite{sciortino3}, where
using a qualitatively
similar interaction potential the authors have found again the behavior
shown in Fig.s \ \ref{figura2} and \ \ref{figura3}, and stressed the role of
percolation. Interestingly they have also shown that this model
is actually able to reproduce the local order
typically observed in the experimental system \cite{bartlett}.

\begin{figure}
\begin{center}
\mbox{ \epsfysize=6cm\epsfbox{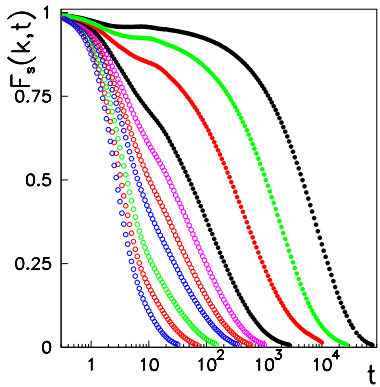}}
\end{center}
\caption{(color online). The incoherent scattering function, $F_s(k,t)$, for
$k_B T=0.25\epsilon$ and $\phi=~0.08$, $0.10$, $0.12$, $0.14$, $0.15$, 
$0.16$, $0.17$,
$0.19$, $0.21$, $0.23$ (from left
to right). Empty circles in figure correspond to values of $\phi
<\phi_p $.}
\label{figure1}
\end{figure}

In order to study the relaxation dynamics, 
we calculate different time autocorrelation functions at equilibrium.
In Fig.~\ref{figure1} the incoherent scattering functions, $F_s(k,t)$, 
are plotted for $k\simeq1.88$, $k_B T=0.25\epsilon$ and
different $\phi$. 
From these functions we extract a characteristic relaxation time,
$\tau_{\alpha}$, defined by $F_s(k,\tau_{\alpha}) = 0.1$, plotted
in Fig.~\ref{figure3}. 

\begin{figure}
\begin{center}
\mbox{ \epsfysize=6cm\epsfbox{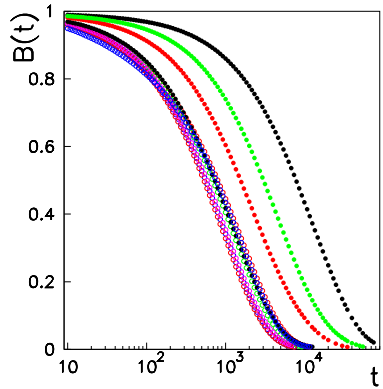}}
\end{center}
\caption{(color online). The bond correlation function, $B(t)$, for
$k_B T=0.25\epsilon$ and $\phi=~0.08$, $0.10$, $0.12$, $0.14$, $0.15$, 
$0.16$, $0.17$,
$0.19$, $0.21$, $0.23$ (from left
to right). Empty circles in figure correspond to values of $\phi
<\phi_p$.}
\label{figure2}
\end{figure}

\begin{figure}
\begin{center}
\mbox{ \epsfysize=6cm\epsfbox{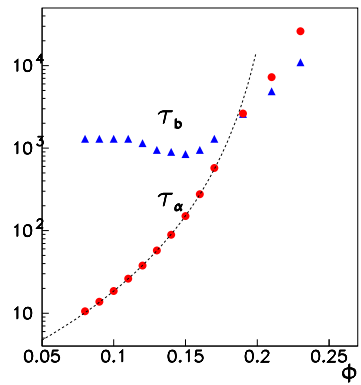}}
\end{center}
\caption{(color online).
The bond lifetime, $\tau_b$,  and
the relaxation time, $\tau_{\alpha}$, 
(in units of MD steps) as a
function of the volume fraction $\phi$ at $k_B T = 0.25 \epsilon$. The curve
in figure is a power law fit, $(\phi_c-\phi)^{-\gamma}$, with
$\phi_c\simeq0.22$ and $\gamma\simeq4.0$.}
\label{figure3}
\end{figure}

We also study the time correlation of the
clusters in terms of the time autocorrelation functions of the bonds,
$B(t)$, which are plotted in Fig.~\ref{figure2}.
\begin{figure}
\begin{center}
\mbox{ \epsfysize=6cm\epsfbox{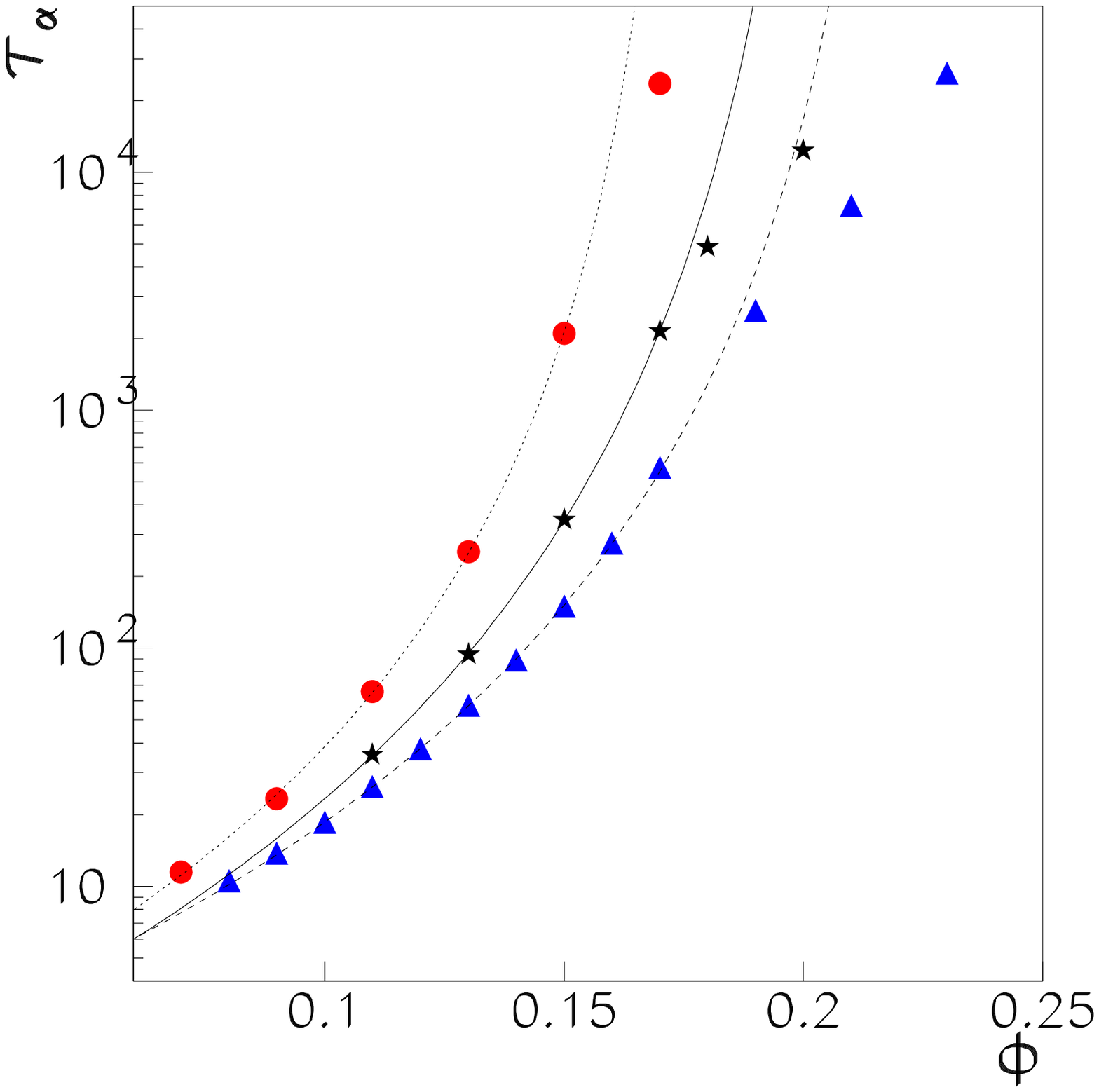}}
\end{center}
\caption{(color online).
The relaxation time, $\tau_\alpha$,
as a function of the volume fraction, $\phi$, for different temperatures, 
$T=~0.25,~0.23,~0.2$ (from right to left).
The curves are power laws, $(\phi_c-\phi)^{-\gamma}$, where respectively
$\phi_c\simeq0.22$ and $\gamma\simeq4.0$, $\phi_c\simeq0.21$ and 
$\gamma\simeq4.3$, and
$\phi_c\simeq0.17$ and $\gamma\simeq3.7$. The percolation threshold is
weakly dependent on the temperature, $0.16\lesssim\phi_p\lesssim 0.17$.}
\label{figure4}
\end{figure}

The function $B(t)$ is defined as
\begin{equation}
B(t)=\frac{\sum_{ij}
\left[ \langle n_{ij}(t)n_{ij}(0)\rangle-\langle n_{ij}\rangle^2\right]}%
{\sum_{ij}\left[\langle n_{ij}\rangle-\langle n_{ij}\rangle^2\right]},
\end{equation}
where $n_{ij}(t)=1$ if particles $i$ and $j$ are linked at time $t$,
$n_{ij}(t)=0$ otherwise. The data show that, differently from the
behavior of $F_s(k,t)$, the bond relaxation does not change much
below the percolation threshold, $\phi_p$, and instead it becomes
slower and slower above the percolation threshold. 
By fitting the curves with stretched exponential functions, we obtain the bond
lifetime $\tau_b$.
In Fig.~\ref{figure3} $\tau_b$ is compared with
$\tau_{\alpha}$. The data show that, at low volume
fractions, the bond lifetime is much larger than the density
relaxation time and in this region $\tau_\alpha$ can be fitted by a
power law.
This result can be understood by considering that
the system dynamically behaves as made of permanent clusters and the
structural relaxation is directly related to the size of the clusters,
like in chemical gelation. At higher volume fraction close to the
percolation threshold, the relaxation time  and the bond lifetime
eventually become of the same order of magnitude and one observes a
deviation from the power law critical behavior. 

Within this scenario, the percolation of long-living structures
is the mechanism which produces the enhancement of the structural relaxation 
time at low volume fraction: Close to the percolation threshold, 
$\tau_\alpha$ increases fast enough to reach the scale of the bond lifetime, 
and the dynamics crosses over to the glassy regime.

In fact in this
region on the time scale of the structural relaxation time, the bonds
cannot be considered as permanent anymore. This explains the
crossover to a glassy regime, in which the structural relaxation time 
depends on the bond finite lifetime and the crowding of the particles. 

At higher temperature when the bond lifetime does not play
any role, structural arrest is only due to the hard sphere component
of the potential.


By decreasing the temperature we find that the
critical regime becomes more extended, due to the increase
of the bond lifetime (see Fig.~\ref{figure4}), and the critical volume fraction
$\phi_c$ tends to the percolation threshold $\phi_p$. Therefore
at very low temperatures we expect the structural
arrest to coincide with the onset of a spanning cluster on the time
scale of the numerical simulations,
in agreement with previous results
\cite{noi3} and recent experiments\cite{bartlett}.

\begin{figure}
\begin{center}
\mbox{ \epsfysize=6cm\epsfbox{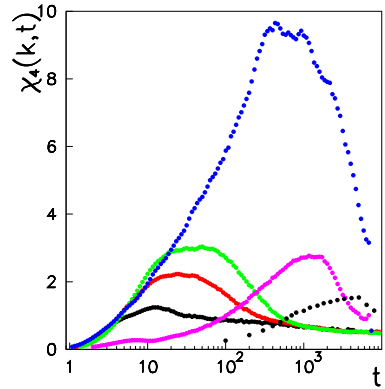}}
\end{center}
\caption{(color online).
The dynamic susceptibility, $\chi_4(k,t)$, for
$k\simeq1.88$,  $k_B T=0.25\epsilon$ and
$\phi=~0.08$, $0.13$, $0.15$, $0.17$, $0.21$, $0.23$ (from left to right).}
\label{figure5}
\end{figure}

\begin{figure}
\begin{center}
\mbox{ \epsfysize=6cm\epsfbox{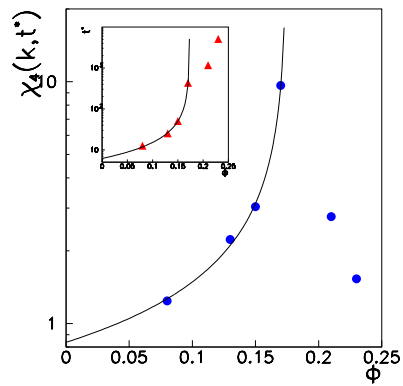}}
\end{center}
\caption{(color online). {\bf Main frame} The maximum value of the dynamic
susceptibility, $\chi_4(k,t^*)$, as a
function of the volume fraction, $\phi$, at $k_B T = 0.25 \epsilon$.
The curve in figure is a power law fit, $(\phi_c-\phi)^{-\gamma}$, with
$\phi_c\simeq0.17$ and $\gamma\simeq0.68$. {\bf Inset} 
The time, $t^*$, (in units of MD steps) as a
function of the volume fraction, $\phi$, at $k_B T = 0.25 \epsilon$.  The curve
in figure is a power law fit, $(\phi_c-\phi)^{-\gamma}$, with
$\phi_c\simeq0.17$ and $\gamma\simeq1.0$.}
\label{figure6}
\end{figure}

In order to further understand and characterize the crossover
region, we have calculated the dynamic susceptibility, defined as the 
fluctuations of $F_s(k,t)$ \cite{toninelli}: 
\be \chi_4(k,t)=N (\lan F_s(k,t)^2\ran -\lan
F_s(k,t)\ran^2). \label{eq_chi} \ee Similar quantities, largely
studied in glassy systems (see Refs.\ \cite{franz, glotzer}), are
related to the presence of dynamic heterogeneities, spatial
correlations between particles that move faster than the average. In
Fig.\ \ref{figure5}, $\chi_4(k,t)$ is plotted as function of $t$, for
$k\simeq1.88$, $k_B T=0.25\epsilon$ and different values of $\phi$.
We find that $\chi_4(k,t)$  exhibits a maximum at time $t^*$, which
first increases and then decreases signalling the presence of a
crossover as clearly shown in Fig.~\ref{figure6}, where the time,
$t^*$, and the maximum value, $\chi_4(k,t^*)$, are plotted as a
function of the volume fraction: At low volume fraction they both
follow a power law diverging,
whereas a deviation from this critical behavior appears at
volume fraction higher than the percolation threshold.


The presence of a maximum at a time $t^*$, increasing together with
the height of the maximum was recently found in experiments on
colloidal gels as colloidal gelation was
approached \cite{cipelletti}.



In conclusions in our model the presence of two well separated time scales
at low temperature and low volume fraction originates a colloidal gelation
related to the formation of persistent structures: Increasing the volume
fraction the two time scales become comparable and a crossover from a gel-like
behavior to a glass-like one is found.

This work has been partially supported by the Marie Curie Fellowship
HPMF-CI2002-01945 and Reintegration Grant MERG-CT-2004-012867,
EU Network Number MRTN-CT-2003-504712, MIUR-PRIN 2004, MIUR-FIRB 2001,
CRdC-AMRA, INFM-PCI.

%
%

%

\begin{thebibliography}{99}
\bibitem{lamellar} The system presents a transition to a
tubular or lamellar phase \cite{tarzia}
at increasing volume fraction. In particular we find that some
configurations at $\phi\ge 0.19$ are in such a phase.
The data here presented are limited to configurations where no ordered phase
is observed.
\bibitem{notahill} At low temperatures, such as the temperatures studied in this
paper, this definition is practically equivalent to the Hill definition
adopted in first paper of Ref.\ \cite{noi3}.
\end{thebibliography}
\end{document}